# Chemical Synthesis of Nano-Sized particles of Lead Oxide and their Characterization Studies

M.Alagar, T.Theivasanthi*, A. Kubera Raja

Center for Research and Post Graduate Department of Physics, Ayya Nadar Janaki Ammal College, Sivakasi – 626124, Tamilnadu, India.

**Abstract**

*The quantum dots of semiconductor display novel and interesting phenomena that have not been in the bulk material. The color tunability is one of the most attractive characteristics in II-VI semiconductor nanoparticles such as CdS, ZnS, CdSe, ZnSe and PbO. In this work, the semiconductor lead oxide nanoparticles are prepared by chemical method. The average particle size, specific surface area, crystallinity index are estimated from XRD analysis. The structural, functional groups and optical characters are analyzed with using of SEM, FTIR and UV- Visible techniques. The optical band gap value has also been determined.*

## 1. Introduction

Nanotechnology involves the study, control and manipulation of materials at the nanoscales, typically having dimensions less than 100nm.(K.Mallikarjuna *et al.,* 2012; Natarajan Kannan and Selvaraj Subbalaxmi, 2011). The properties of such materials are novel and can be engineered by controlling the dimensions of these building blocks and their assembly via physical, chemical or biological methods. (Manoj Singh *et al.,* 2011). The characters of metal and metal oxide nanoparticles like optical, electronic, magnetic, and catalytic are depending on their size, shape and chemical surroundings. (Theivasanthi and Alagar 2010).

This paper is discussing about easy, simple, fast and low cost preparation i.e. chemical synthesis of lead oxide nanoparticles and its characterizations - XRD, SEM, FT-IR and UV-Vis.(Bhupinder Singh Sekhon, 2012). Chemical method can be used to prepare wide range of materials. (Das *et al.*, 2009; K. Satyavani *et al.*, 2011).

___________________________________________
**\*Corresponding author.**     *E-mail*:  theivasanthi@pacrpoly.org

## 2. Experimental Method

Lead Oxide (PbO) semiconductor nanoparticles were prepared by Chemical synthesis method. 60 ml of 1.0 M Pb($C_2H_3O_2$)$_2$.3 $H_2O$ (lead (II) acetate) aqueous solution was prepared using de-ionized water and heated upto 90 $^o$C. This solution was added to an aqueous solution of 50 ml of 19M NaOH in a beaker and stirred vigorously. Upon adding the lead (II) acetate, the solution initially became cloudy, and then turned a peach colour, and finally a deep orange red. At this position, stirring was stopped, and the precipitate was allowed to settle. The supernatant was then decanted, filtered on a Buchner funnel, washed with de-ionized water repeatedly, and dried for overnight in a drying oven at 90 $^o$C. The sample was then removed and lightly crushed in a mortar and pestle. Its structural characterizations were done for confirmation of lead oxide nanoparticles.

XRD analysis of the prepared sample was done using a X'pert PRO of PANalytical diffractometer, Cu-Kα X-rays of wavelength (λ)=1.54056 Å and data was taken for the 2θ range of 10° to 90° with a step of 0.02°. Particle size, specific surface area and crystallinity Index of the particles were derived from this analysis reports. The surface morphology was analyzed by using SEM Model S-3000H of HITACHI. Functional groups were analyzed by SHIMADZU – 8400S FT-IR spectrometer. The FTIR spectra were recorded for the range of 400 cm$^{-1}$ to 4000 cm$^{-1}$. Optical studies were done with SHIMADZU UV-300 and the measurements were done for the wavelength range of 190 nm to 800 nm.

## 3. Results and Discussion

### 3.1. XRD – Particle Size

Figure 1 shows the XRD pattern of Lead Oxide nanoparticles. The average grain size of Lead Oxide nanoparticles is determined using *Debye-Scherrer formula*. (Nath *et al*., 2007;

P.K. Dash and Y. Balto, 2011; M. Baneto *et al.*, 2010).

$$D = \frac{0.9\lambda}{\beta\cos\theta} \quad (1)$$

Where 'λ' is wave length of X-Ray (0.1541 nm), 'β' is FWHM (full width at half maximum), 'θ' is the diffraction angle and 'D' is particle diameter size. The particle size is calculated using above formula is 60 nm.

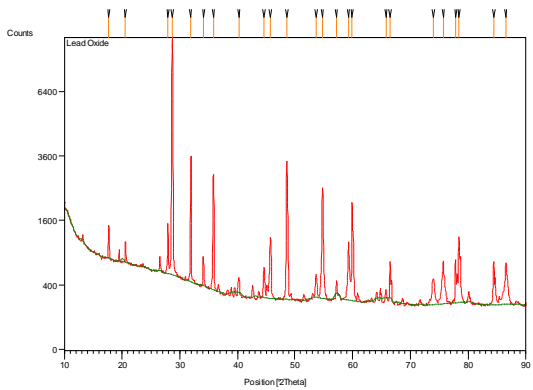 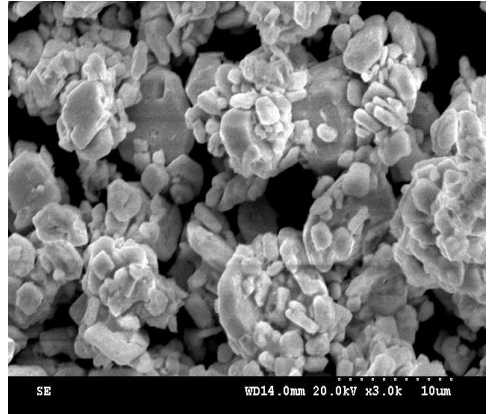

Figure 1. XRD Spectrum of Lead oxide (Pbo) nanoparticles.  Figure 2. SEM picture showing Lead oxide nanoparticles.

## 3.2. XRD - Specific Surface Area

Specific surface area (SSA) is a derived scientific value that can be used to determine the type and properties of a material. It has a particular importance in case of adsorption, heterogeneous catalysis and reactions on surfaces. SSA is the SA per mass.

$$SSA = \frac{SA_{part}}{Vpart * density} \quad (2)$$

Here SSA is Specific surface area, SApart is Surface Area of particle, Vpart is particle volume and density is lead oxide density value. (Jiji Antony *et al.*, 2006).

$$S = 6 * 10^3 / D_p \rho \tag{3}$$

Where S is the specific surface area, Dp is the size of the particles, and ρ is the density of lead oxide nanoparticles. (J.Y.Park *et al.*, 2006). Mathematically, SSA can be calculated using these formulas 2 and 3. SSA calculation details of sample are presented in Table.1.

Table 1. Specific Surface Area of Lead oxide Nanoparticles

| Particle Size (nm) | Surface Area ($nm^2$) | Volume ($nm^3$) | Density ($g/cm^3$) | SSA ($m^2/g$) | SA to Volume Ratio |
|---|---|---|---|---|---|
| 60 | 11304 | 113040 | 9.5 | 10.52 | 0.1 |

### 3.3. XRD - Crystallinity Index

Sharper XRD peaks are typically indicative of high nanocrystalline nature. Crystallinity is evaluated through comparison of particle size determined by Scherrer equation with size ascertained by SEM. Crystallinity index Eq. is below.

$$I_{cry} = \frac{D_p(SEM, TEM)}{D_{cry}(XRD)} (I_{cry} \geq 1.00) \tag{4}$$

Where $I_{cry}$ is the crystallinity index; $D_p$ is the particle size (obtained from either TEM or SEM); $D_{cry}$ is the particle size (calculated from XRD). If $I_{cry}$ value is close to 1, then it is assumed that the crystallite size represents monocrystalline whereas a polycrystalline have a much larger crystallinity index. (X.Pan *et al.*, 2010). Table.2. displays the crystallinity index of the sample that scored higher than 1.0 which indicates crystalline nature of sample.

Table 2. The Crystallinity Index of Lead oxide Nanoparticles.

| Sample | Dp (nm) | Dcry (nm) | Icry (unitless) | Particle Type |
|---|---|---|---|---|
| Lead oxide Nanoparticles | 99 | 60 | ~1.65 | Polycrystalline |

### 3.4. SEM studies

The results of SEM morphological and nanostructural studies of the Lead Oxide nanopartcles are shown in Figure 2.

### 3.5. FTIR studies

FTIR spectra for the Lead Oxide Nanoparticles are shown in the Fig.3. The absorption peak at 466.74 cm$^{-1}$ indicates the presence of Pb-O Stretching and also the peak at 557.39 cm$^{-1}$ indicates the presence of oxides. These two peaks are very sharp. It is confirmed that the final product is the presence of lead and oxide.

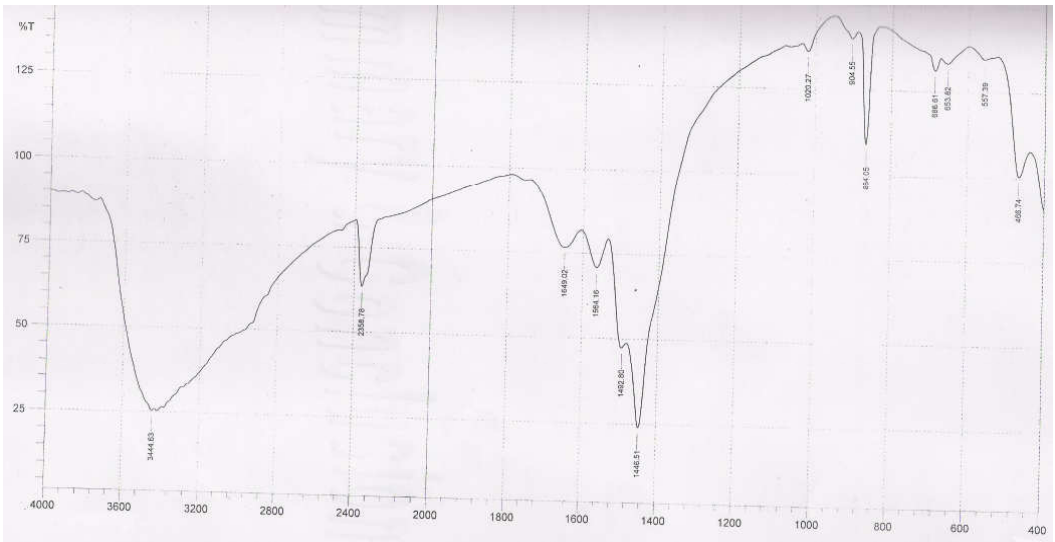

Figure 3. FTIR Spectrum of PbO Nanoparticles.

### 3.6. UV-Visible studies

The UV-study of the samples was carried out to explore their optical properties. The UV-Visible spectrum and energy band gap spectrum are shown in Figure 4 and Figure 5 respectively. The band gap value of the sample is 3.82 eV. The energy band gap value of the prepared sample is estimated using Tauc relation. (Rema Devi *et al*., 2007)

$$\alpha h\upsilon = A(h\upsilon - E_g)^n \tag{5}$$

Where, A is Absorption Coefficient, $h\nu$ – Photon energy, Eg – Energy band gap, n = ½ for the direct band gap transition, n = 2 for the indirect band gap transition.

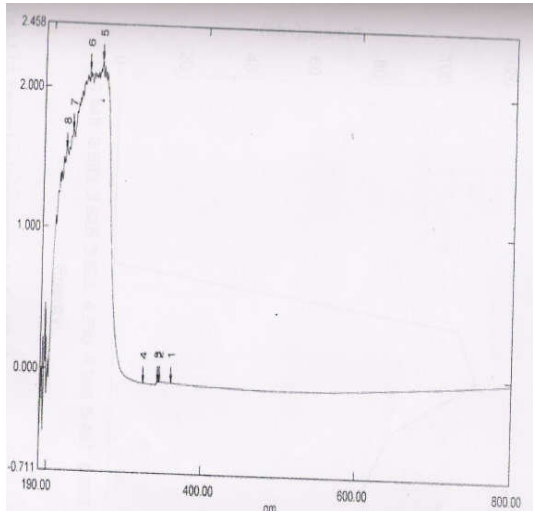

Figure 4. UV-Visible Spectrum of PbO nanoparticles.

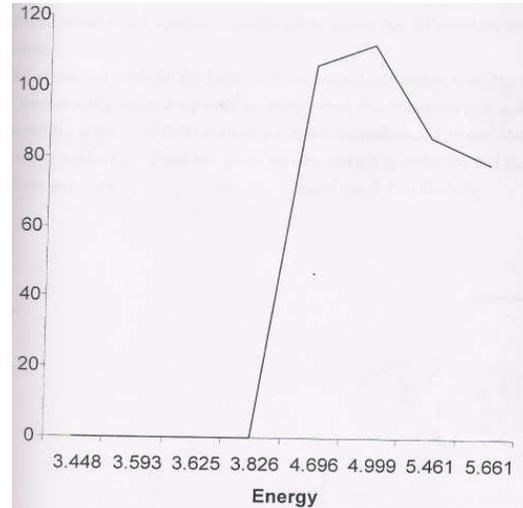

Figure 5. Energy Band gap Spectrum of PbO nanoparticles.

The shift in band gap nanoparticle is the due to the quantum confinement and it has the quantitative form. (Manickathai *et al.*, 2008).

$$\Delta E_g = E_g^{nano} - E_g^{bulk} = h^2 \pi^2 / 2MR^2 \qquad (6)$$

Where, R is Radius of the particles and M-effective mass of the electron.

## 4. Conclusions

Lead (II) oxide nanoparticles have been synthesized successfully in rapid and easy chemical method. XRD, SEM, FTIR and UV- Visible characterizations techniques confirm the results. Litharge has been developed which results in the formation of high-purity nanoparticles. The experimental scheme produces lead oxide nanoparticles with no detectable amounts of yellow massicot form.

## Acknowledgements

The authors thanks to staff & management of *PACR Polytechnic College*, Rajapalayam, India and *Ayya Nadar Janaki Ammal College*, Sivakasi, India for their support.